\documentclass[11pt,a4paper]{article}
\pdfoutput=1
\usepackage{jheppub}
\usepackage{graphicx}
\usepackage{slashed}
\newcommand{\ba}{\begin{eqnarray}}
\newcommand{\ea}{\end{eqnarray}}
\newcommand{\no}{\nonumber}
\title{
A dangerous irrelevant UV-completion of the composite Higgs
}
\date{\today
}
\author{Luca Vecchi}
\affiliation{Maryland Center for Fundamental Physics,\\ Department of Physics, University of Maryland\\
College Park, MD 20742, USA}
\emailAdd{vecchi@umd.edu}
\abstract{
One of the most challenging hurdles to the construction of realistic composite Higgs models is the generation of Yukawa couplings for the Standard Model fermions. This problem can be successfully addressed in approximate conformal theories that admit a marginally relevant mixing between composite fermionic operators and the SM fermions. I argue that SU(3) gauge theories with light Dirac flavors in the fundamental representation feature all the ingredients under theoretical control, including a strongly-coupled IR fixed point, composite partners for all Standard Model fermions, absence of Landau poles at low energy, and a realistic phenomenology. These models acquire the status of compelling UV-completions of the SM if some spin-1/2 baryon operator has scaling dimension close to 2.5 within the conformal window, a possibility that can only be assessed via non-perturbative methods like lattice QCD. A distinctive collider signature is long-lived hadrons with fractional charges.

Vacuum alignment is controlled by the Nambu-Goldstone bosons of the coset $SU(4)\times SU(4)/SU(4)$. With a technically natural choice of mixing for the top-quark, the exotic scalars with electro-weak charges acquire large positive masses and a compelling custodial-symmetric phenomenology is obtained. In the decoupling limit the symmetry breaking pattern effectively reduces to $SU(4)\to Sp(4)$ with a light Higgs.
}
\begin{document}
\maketitle
%%%%%%%%%%%%%%%%%%%%%%%%%%%%%%%%%%%%%%%%%%%%%%%%%%%%%%%%%%%%%%%%%%%
%
%%%%%%%%%%%%%%%%%%%%%%%%%%

\section{Introduction}

Solutions of the hierarchy problem that postulate a new strong dynamics at the TeV scale must deal with three difficult tasks: suppress contributions to the electro-weak parameters, account for a light Higgs boson with a mass of $126$ GeV and couplings similar to those predicted by the Standard Model (SM), and satisfy flavor constraints. 

To account for the existence of a {\emph{parametrically}} light composite Higgs boson we require that the strong dynamics has a global symmetry ${\cal G}$ spontaneously broken at a scale $\Lambda\sim4\pi f$ to a subgroup ${\cal H}$ weakly-gauged to the electro-weak symmetry. ${\cal G}/{\cal H}$ must contain an electro-weak doublet Nambu-Goldstone boson (NGB), and the explicit ${\cal G}$ breaking should align the vacuum such that $v^2/f^2<1$, where $v\simeq246$ GeV~\cite{Kaplan:1983fs}. A custodial symmetry $\subset {\cal H}$ may be invoked to protect the $T$ parameter from large corrections, and $\Lambda\gtrsim$ few TeV allows us to sufficiently suppress the contributions to other electro-weak observables.  

Under the above conditions the couplings of the NGB Higgs approach those of an elementary scalar as $v^2/f^2\to0$~\cite{Giudice:2007fh}. Similarly, flavor violation beyond the SM can be (at least morally) decoupled in this limit.

The real challenge that any strong Higgs dynamics faces is that it is not obvious how to couple the composite sector to the SM fermions to generate a Yukawa coupling of order unity for the top quark. If we assume the SM fermions $Q$ are elementary fields, external to the strong dynamics, we have two options: either we couple the strong sector to SM fermion bilinears or linearly to $Q$. In the first option the Yukawas arise from a bilinear interaction of the form $QQ\Phi$, with $\Phi$ a scalar operator of the strong dynamics (usually a fermion bilinear). The problem with this picture is that the Yukawa couplings are always irrelevant, and therefore suppressed by powers of $\Lambda/\Lambda_{\rm F}$, where $\Lambda_{\rm F}$ is the ``flavor scale" at which the above operator is generated. Because $\Lambda_{\rm F}$ plays the role of a cutoff for the strong Higgs dynamics, this means that our description has a very limited range of validity.

The alternative is to consider interactions of the form
\ba\label{PCi}
\lambda QO,
\ea
provided color $SU(3)_c$ is a subgroup of ${\cal H}$ and $O$ is a fermionic operator of the strong dynamics. At the confinement scale $O$ interpolates a resonance $\Psi$ (we normalize $O$ according to $O\to\frac{\Lambda}{g_*}\Psi$) and (\ref{PCi}) describes a mixing of order $\lambda/g_*$ between $\Psi$ and the elementary fermion. The observed SM fermions are a mixture of both $Q$ and $\Psi$, they are {\emph{partially composite}}~\cite{Kaplan:1991dc}. For this reason this approach is known as Partial Compositeness (PC).~\footnote{For a review see \cite{Contino:2010rs} and references therein.}

Including the renormalization group (RG) evolution from $\Lambda_{\rm F}$ down to the confinement scale the SM fermion Yukawa will formally read
\ba\label{yt}
y=c\frac{Y_*}{g_*^2}\lambda_{L}(\Lambda)\lambda_{R}(\Lambda),
\ea
with $c$ a matrix in flavor space with entries of order unity and $Y_*$ measuring the strength of the coupling between the fermionic resonances and the Higgs at the scale $\sim\Lambda$. In a generic small $N$ strong dynamics $Y_*\sim g_*\sim4\pi$. The alternative (\ref{PCi}) was first pointed out by D. B. Kaplan in a Technicolor framework~\cite{Kaplan:1991dc} and is effectively realized in Randall-Sundrum models with fermions in the bulk (and the Higgs an accidentally light resonance~\cite{Gherghetta:2000qt}\cite{Huber:2000ie}, a NGB~\cite{Contino:2003ve}\cite{Agashe:2004rs}, or absent~\cite{Cacciapaglia:2004rb}).

When $\lambda$ is irrelevant we find ourselves in a situation similar to the coupling $QQ\Phi$. 

On the other hand, a qualitatively different picture is possible, at least in principle: (\ref{PCi}) may be a {\emph{marginally relevant}} operator without necessarily implying the existence of strongly relevant operators in the Lagrangian. To see this we assume the strong dynamics is nearly conformal below $\Lambda_{\rm CW}<\Lambda_{\rm F}$, so $O$ should be viewed as a fermionic operator of a conformal field theory (CFT) with a well defined scaling dimension $d_O$.~\footnote{The requirement that the dynamics is an approximate CFT in the range $\Lambda<\mu<\Lambda_{\rm CW}$ is not just needed to unambiguously define $d_O$. The main reason is that the desired hierarchy $\Lambda\ll\Lambda_{\rm F}$ can only be achieved {\emph{naturally}} if all couplings are nearly marginal, i.e. if they are close to a fixed point.} To get a marginally relevant $\lambda$ it is sufficient that $d_O\lesssim2.5$, a requirement that is safely above the unitarity bound $d_O\geq1.5$. Furthermore, up to small violations of factorization, the singlet $\overline{O}O$ (if allowed by the Lorentz symmetry) is expected to be irrelevant for $d_O>2$. Hence, in the interesting regime $d_O\lesssim2.5$ the mass scale $\Lambda_{\rm F}$ associated to the dynamics generating (\ref{PCi}) may be safely decoupled from the TeV, and perhaps taken as large as the Planck scale. In this case the major hurdle to the construction of viable composite Higgs models is overcome.

Kaplan's original goal was to show that $\lambda$ can generate an interesting and novel flavor structure. This is even more so when the dynamics of $O$ is nearly conformal. Indeed, the running of the parameters $\lambda$ in this case depends significantly on the scaling dimension of the associated operator $O$, so even a modest flavor-dependence of the $d_O$s may naturally result in a hierarchical structure of the IR observable (\ref{yt}). In this picture the SM mass hierarchy may thus be generated via RG evolution. The possibility of explaining the flavor puzzle via couplings to a CFT has been appreciated in Randall-Sundrum scenarios with fermions in the bulk~\cite{Gherghetta:2000qt}\cite{Huber:2000ie} as well as in Supersymmetric solutions of the hierarchy problem~\cite{Nelson:2000sn}.

To summarize, (\ref{PCi}) is the only possible coupling between elementary fermions $Q$ and the strong Higgs dynamics that may allow us to construct extensions of the SM without elementary scalars valid all the way to the Planck scale. As a bonus, the same picture can potentially explain the SM flavor hierarchy via RG effects. 

Crucially, to exploit the full potential of PC, 
\begin{itemize}
\item[${\color{red}\circledast}$] the strong dynamics must have an approximate conformal fixed point where the $O$s have scaling dimensions $d_O\lesssim2.5$.
\end{itemize}
Theories with weakly-coupled scalars can easily satisfy this latter requirement (to avoid a fine-tuning of the mass terms one may invoke Supersymmetry, see~\cite{Caracciolo:2012je}\cite{Marzocca:2013fza}\cite{Parolini:2014rza}). The key hypothesis ${\color{red}\circledast}$ is highly non-perturbative in theories without elementary scalars. In this latter case it is not clear how PC is realized in the microscopic theory. One possible approach has been taken in~\cite{Barnard:2013zea} (see also \cite{vonGersdorff:2015fta}), where a Nambu--Jona-Lasinio dynamics was used to model the non-perturbative theory. Another possibility is to use lattice techniques to test whether non-abelian gauge theories in the conformal window can realize ${\color{red}\circledast}$. This program requires explicit candidate UV completions; however not much literature exists on this topic. One reason is that the particle physics community mostly focused on the collider phenomenology of PC, for which an effective description is enough. The relevant dynamics is a strongly-coupled CFT below a very large scale $\Lambda_{\rm CW}$, and in the energy range $\Lambda<\mu<\Lambda_{\rm CW}$ warped extra dimensions with fermions in the bulk provide a very efficient formulation~\cite{Contino:2003ve}\cite{Agashe:2004rs}.

To identify candidate UV completions within non-abelian gauge theories we may systematically look for scenarios that satisfy the requirements that are well under theoretical control, and then leave the harder test of ${\color{red}\circledast}$ to the lattice. In~\cite{Ferretti:2013kya} simple models with cosets $SU(N_F)/Sp(N_F)$ and $SU(N_F)/SO(N_F)$ were identified (see also~\cite{Ferretti:2014qta}\cite{Golterman:2015zwa} for a related study). These authors however did not demand a number of constraints that are essential to decouple $\Lambda_{\rm F}$ and thus realize the ambitious picture described above, specifically the presence of a strong IR fixed point, the absence of Landau poles at low energy, and the possibility of finding partners $O$ for {\emph{all}} SM fermions. The last requirement becomes necessary if the flavor scale is to be truly decoupled from $\Lambda$.

Here I introduce a class of $SU(3)$ gauge models and $N_F$ light flavors with symmetry breaking pattern $SU(N'_F)^2\to SU(N'_F)$, where $N'_F<N_F$ (see Section~\ref{sec:model}). These models have a NGB Higgs, a custodial symmetry, fermionic partners $O$ for all SM fermions, automatic suppression of proton decay, no Landau poles below the Planck scale, and a strongly-coupled IR fixed point broken naturally at $\Lambda\ll\Lambda_{\rm CW}$. Their UV cutoff $\Lambda_{\rm F}$ can thus be consistently taken to be much larger than $\Lambda$, or even sent to the Planck scale, if ${\color{red}\circledast}$ is satisfied.

For $\lambda$ sufficiently small in the UV and $d_O$ sufficiently close to $2.5$ I find that the RG evolution of the coupling (\ref{PCi}) can be determined in generality (see Appendix~\ref{app:IRFP}). The renormalized coupling agrees with that found in 5D models of PC. In the present context this result can be used to argue that (\ref{PCi}) is a small perturbation of the nearly-conformal non-abelian theory. 

A generic implication of PC is the presence of colored scalars at the TeV. Amusingly, QCD-like realizations predict that the lightest color triplet is collider-stable, suggesting novel signatures of PC (see Section~\ref{sec:pheno}). Importantly, a realistic vacuum alignment and Higgs potential can be obtained, as discussed below. I end with a few comments in Section~\ref{sec:conclusions}.

\section{A QCD-like description of PC}
\label{sec:model}

The operator $O$ in (\ref{PCi}) may in principle be $G^a_{\mu\nu}\sigma^{\mu\nu}\psi^a$, where $\psi^a$ is an adjoint fermion of the strong dynamics and $G^a_{\mu\nu}$ the field strength of the strong gauge vector (see also \cite{Ferretti:2013kya}), a 3-fermion operator, or an operator of larger engineering dimension. In view of the fact that $d_O\sim2.5$ is preferred, the former option is the most attractive. However, for any gauge group in this case the non-abelian theory becomes IR-free as soon as we demand a partner for $t_{R}$ and one quark doublet ($\supset t_L$). Hence, with this option all fermion masses are in practice controlled by other interactions; as a result, $O=G^a_{\mu\nu}\sigma^{\mu\nu}\psi^a$ by itself is not enough to construct a realistic model of PC. We will therefore focus on 3-fermion operators. Examples with coset $SU(N_F)/SO(N_F)$ and $SU(N_F)/Sp(N_F)$ have been discussed in~\cite{Ferretti:2013kya}. Here I will present models based on $SU(N_F)^2/SU(N_F)$.

I consider the fermions shown in table~\ref{m1}, where $a$ is a real parameter discussed below.~\footnote{It is possible to embed the fields of table~\ref{m1} into (in)complete representations of a simple group. The key observation is that $S,S'$ form an $SU(2)_{\rm cust}$ doublet with the same hyper-charge as $D$. Strictly speaking, $S'$ is not necessary to generate the SM quark masses: it has been introduced with the sole objective of preserving a custodial symmetry in the IR.} Switching off the SM gauge couplings our models reduce to a strong $SU(3)$ gauge dynamics with $N_F$ (Dirac) flavors in the fundamental representation, where part of the vector subgroup is weakly gauged to the SM $SU(3)_c\times SU(2)_w\times U(1)_Y$. The actual number of constituents $N_F$ depends on which fields have a family index (see below) and how the lepton partners are implemented. As a prototypical realization, consider for instance a model with $a=0$ and $S$ replicated in 3 families; this has $N_F=9$ and passes all the requirements analyzed in the following.

%%%%%%%%%%%%%%%%%%%%%
\begin{table}[h]
\begin{center}
\begin{tabular}{c|cccc} 
\rule{0pt}{1.2em}%
$$ & $SU(3)$ &  $SU(3)_c$ & $SU(2)_w$ & $U(1)_Y$ \\
%\hline
\hline
$T$ & ${\bf 3}$ & ${\bf 3}$ & ${\bf 1}$ & $a$ \\
$D$ & ${\bf 3}$ & ${\bf 1}$ & ${\bf 2}$ & $\frac{1}{3}-\frac{1}{2}a$ \\
$S$ & ${\bf 3}$ & ${\bf 1}$ & ${\bf 1}$ & $-\frac{1}{6}-\frac{1}{2}a$ \\
$S'$ & ${\bf 3}$ & ${\bf 1}$ & ${\bf 1}$ & $\frac{5}{6}-\frac{1}{2}a$ \\
\end{tabular}
\end{center}
\caption{\small Quantum numbers of the constituents $\psi$. The conjugates $\overline{\psi}$ are not shown for brevity. The charges of the SM fermions under $SU(3)_c\times SU(2)_w\times U(1)_Y$ are taken to be $q\sim(3,2)_{1/6}, u\sim(\overline{3},1)_{-2/3}, d\sim(\overline{3},1)_{1/3}, \ell\sim(1,2)_{-1/2}, e\sim(1,1)_{1}$.
\label{m1}}
\end{table}
%%%%%%%%%%%%%%%%%%%%%%%

\subsection{The UV theory}
\label{sec:UV}

We are now interested in the Lagrangian at the UV cutoff $\Lambda_{\rm F}$. I will focus on the quark sector, but will comment on the lepton fields shortly.

Up to operators of dimension $6$ the Lagrangian reads
\ba\label{EFT}
{\cal L}_{\rm UV}(\mu<\Lambda_{\rm F})={\cal L}_{\rm Kin}+{\cal L}_{\rm mass}+{\cal L}_{\rm self}+{\cal L}_{\rm gauge}+{\cal L}_{\rm SM,EFT}+{\cal L}_{\rm PC}+{\cal L}_{\rm ETC}.
\ea
Interactions of dimension $>6$ are assumed to be irrelevant and will be ignored. 

At dimension $\leq4$ the only operators allowed by the gauge symmetries are the kinetic terms for the SM fields (minus the Higgs) and those in table \ref{m1}, included in ${\cal L}_{\rm Kin}$, plus mass terms for the exotic fields:
\ba\label{ma}
{\cal L}_{\rm mass}=-m_TT\overline{T}-m_DD\overline{D}-m_SS\overline{S}-m_{S'}S'\overline{S'}+{\rm hc}.
\ea
As usual, these masses can naturally be much smaller than $\Lambda_{\rm F}$ because of an approximate axial symmetry.

At the ``non-renormalizable" level we find 
\ba\label{self}
{\cal L}_{\rm self}=\frac{C_{\psi5}}{\Lambda_{\rm F}}\overline{\psi} T^a\sigma_{\mu\nu}\psi G^a_{\mu\nu}+{\rm hc}+\frac{C_{G}}{\Lambda^2_{\rm F}}GGG+{\cal L}_{4\psi},
\ea
where $\psi=T,D,S,S'$. Next, there are dimension-5 dipole couplings to the SM gauge fields of the form $\overline{\psi}\sigma_{\mu\nu}\psi F_{\mu\nu}\subset{\cal L}_{\rm gauge}$. Most of these operators are suppressed by the same axial symmetry weakly broken by (\ref{ma}), with the exception of $GGG$, $|\psi\psi'|^2$, and $|\psi\overline{\psi'}|^2$.

In addition there is a long list of operators constructed with SM fields only, ${\cal L}_{\rm SM,EFT}$.

The interesting piece of the UV Lagrangian involves both $\psi$ and the SM fermions. The Higgs constituents have the right quantum numbers to couple linearly to all SM quark representations via PC interactions of the form (\ref{PCi}). For any value of the hyper-charge parameter $a$ we can write:
\ba\label{PC}
{\cal L}_{\rm PC}&=&\frac{C_q}{\Lambda_{\rm F}^2}q\overline{TDS}+\frac{C_u}{\Lambda_{\rm F}^2}uTDD+\frac{C_u'}{\Lambda_{\rm F}^2}uTSS'+\frac{C_d}{\Lambda_{\rm F}^2}dTSS+{\rm hc}.
\ea
To save typing we used a symbolic notation for the baryonic operators $O_{q,u,d}$. Explicit expressions are given in table~\ref{t}. Note that (\ref{PC}) may be allowed by the approximate axial symmetry if the SM fermions are assigned appropriate charges~\cite{Kaplan:1991dc}.

%%%%%%%%%%%%%%%%%%%%%
\begin{table}[h]
\begin{center}
\begin{tabular}{c|cc|c} 
\rule{0pt}{1.2em}%
$(O_q)_I$ & $(O_u)$ &  $(O_u')_I$ & $(O_d)_{IJ}$ \\
\hline
\hline
$\overline{T}(\overline{DS_I})$ & $D(TD)$ & $T(S_IS')$ & $T(S_{I}S_{J})$\\
$\overline{D}(\overline{TS_I})$ & $$ & $S_I(TS')$ & $S_I(TS_J)$\\
\hline
$\overline{T}({DS_I})^\dagger$ & $D(\overline{TD})^\dagger$ & $T(\overline{S_IS'})^\dagger$ & $T(\overline{S_{I}S_{J}})^\dagger$\\
$\overline{D}({TS_I})^\dagger$ & $$ & $S_I(\overline{TS'})^\dagger$ & $S_I(\overline{TS_J})^\dagger$\\
$\overline{S_I}({TD})^\dagger$ & $$ & $S'(\overline{TS_I})^\dagger$ & $$\\
%\hline
%$({T}^\dagger\bar\sigma^\mu\overline{S}) D^\dagger\bar\sigma_\mu$ & $(\overline{D}^\dagger\bar\sigma^\mu D)\overline{T}^\dagger\bar\sigma_\mu$ & $(\overline{T}^\dagger\bar\sigma^\mu S_I)\overline{S'}^\dagger\bar\sigma_\mu$ & $(\overline{T}^\dagger\bar\sigma^\mu S_I)\overline{S_J}^\dagger\bar\sigma_\mu$\\
%$({T}^\dagger\bar\sigma^\mu\overline{D}) S^\dagger\bar\sigma_\mu$ & $(\overline{D}^\dagger\bar\sigma^\mu T)\overline{D}^\dagger\bar\sigma_\mu$ & $(\overline{T}^\dagger\bar\sigma^\mu S')\overline{S_I}^\dagger\bar\sigma_\mu$ & $(\overline{S_I}^\dagger\bar\sigma^\mu T)\overline{S_J}^\dagger\bar\sigma_\mu$\\
%$({S}^\dagger\bar\sigma^\mu\overline{T}) D^\dagger\bar\sigma_\mu$ & $$ & $(\overline{S_I}^\dagger\bar\sigma^\mu T)\overline{S'}^\dagger\bar\sigma_\mu$ & $$\\
\end{tabular}
\end{center}
\caption{\small Spin-$1/2$ partners of the SM fermions $q,u,d$ of (\ref{PC}) in a scenario where $S$ appears in $3$ flavors ($I, J=1,2,3$) and $a$ is arbitrary. The Fierz identities have been used to eliminate redundant expressions. It is understood that the $SU(3)$ indices are contracted via the fully anti-symmetric tensor, whereas the round parenthesis denote contractions of the Lorentz indices.  \label{t}}
\end{table}
%%%%%%%%%%%%%%%%%%%%%%%

Finally, we have interactions of the form $QQ\Phi$: 
\ba\label{ETC}
{\cal L}_{\rm ETC}&=&\frac{C_{qu}}{\Lambda_{\rm F}^2}quD\overline{S}+\frac{C'_{qu}}{\Lambda_{\rm F}^2}qu\overline{D}S'+\frac{C_{qd}}{\Lambda_{\rm F}^2}qd\overline{D}S+\frac{C'_{qd}}{\Lambda_{\rm F}^2}qdD\overline{S'}\\\no
&+&\frac{C_{\ell e}}{\Lambda_{\rm F}^2}\ell e\overline{D}S+\frac{C'_{\ell e}}{\Lambda_{\rm F}^2}\ell eD\overline{S'}+\frac{C_{Q^\dagger Q}}{\Lambda_{\rm F}^2}Q^\dagger\overline{\sigma}^\mu Q\psi^\dagger\overline{\sigma}_\mu\psi+{\rm hc},
\ea
with $Q=q,u,d,\ell,e$. These latter are irrelevant, and some also suppressed by the approximate axial symmetry.

\paragraph{Flavor indices}

To allow a mixing between all generations of $q,u,d$ and composite operators $O_{q,u,d}$ of a given $SU(N_F)\times SU(N_F)$ representation (see table~\ref{t}), I introduce a family index in some of the fields in table~\ref{m1}. There are several ways this can be done. In the following we will focus on a scenario in which $S$ appears in 3 families, which is the one with the smallest number of flavors ($N_F=9$).

 \paragraph{Accidental symmetries}

For some $a$ the UV Lagrangian also contains lepto-quark operators of the form $\ell q\overline{\psi}\psi'$. On the other hand, I find that for
\ba\label{aa}
a\neq\pm1,\pm\frac{1}{3}, \frac{5}{3},\frac{7}{3},-\frac{11}{9},-\frac{5}{9},\frac{1}{9},\frac{7}{9},\frac{13}{9},\frac{15}{9},
\ea
there exist no operator in ${\cal L}_{\rm UV}-{\cal L}_{\rm SM,EFT}$ that mediates proton decay. In fact, those interactions enjoy 3 accidental symmetries (see table \ref{t2}): the usual baryon and lepton numbers and a ``techni-family" number $U(1)_{\rm T}$. Obviously, $U(1)_{B, L}$ are violated by ${\cal L}_{\rm SM, EFT}$, but this effect can be neglected if $\Lambda_{\rm F}$ is much larger than the TeV.

%%%%%%%%%%%%%%%%%%%%%
\begin{table}[h]
\begin{center}
\begin{tabular}{c|ccc} 
\rule{0pt}{1.2em}%
$$ & $U(1)_B$ &  $U(1)_L$ & $U(1)_{\rm T}$ \\
%\hline
\hline
$q, u^\dagger, d^\dagger$ & $\frac{1}{3}$ & $0$ & $0$  \\
$\ell, e^\dagger$ & $0$ & $1$ & $0$  \\
$T$ & $\frac{1}{3}$ & $0$ & $-2/3$  \\
$D,S,S'$ & $0$ & $0$ & $1/3$  \\
\end{tabular}
\end{center}
\caption{\small Accidental symmetries of ${\cal L}_{\rm UV}-{\cal L}_{\rm SM,EFT}$. The $\overline{\psi}$s have conjugate charges. \label{t2}}
\end{table}
%%%%%%%%%%%%%%%%%%%%%%%

\paragraph{Landau pole for $U(1)_Y$}

Within the perturbative regime some of these QCD-like scenarios can consistently be extrapolated up to the Planck scale. For instance, in the minimal $N_F=9$ model mentioned above the gauge group $SU(3)\times SU(3)_c\times SU(2)_w$ is asymptotically free. As in the SM, hyper-charge has a Landau pole, whose value depends on the parameter $a$. Subtracting the Higgs doublet contribution to the SM and adding that of the fields in table~\ref{m1}, assuming 3 families of $S$, the coefficient $\beta_1'$ for the one-loop beta-function of $g'$ becomes $\beta_1'=32/3 + 2 a(9a-2)$. Requiring that the Landau pole be above the Planck scale we find the condition $-0.5\lesssim a\lesssim0.8$, with the highest value ($\sim10^{28}$ TeV) at $a\sim0$. The choices $a=0,\pm\frac{1}{6},\frac{2}{3},$ etc. are among the simplest options. Similar results are found in other realizations of our model.

When $SU(3)$ becomes strong, at scales smaller than $\Lambda_{\rm CW}$, the new physics contributions to the SM beta functions are expected to slightly depart from the perturbative estimates. However, given that at least for $SU(3)_c\times SU(2)_w$ the latter are comfortably within the asymptotically free regime, I find it reasonable to expect that my conclusion does not change qualitatively at strong coupling.

 \paragraph{The SM leptons} Linear couplings to the SM leptons can be arranged in several ways. One option is to proceed analogously to quarks and introduce additional flavors. We can do this without violating the accidental symmetries in table \ref{t2}. For instance we may add 2 flavors with SM charges $({1},1)_{-5/12+a/4}$, but other options are allowed. This possibility obviously increases $N_F$ and is expected to make the $SU(3)$ coupling weaker.  
 
A second option is to introduce a new strong $SU(3)'$, thus obtaining a special realization of 2-Higgs doublet models with one doublet coupled to quarks and a separate one to leptons. One should then make sure that the $SU(3)'$ confines at a scale comparable to $\Lambda$, which we think is plausible with some model-building. Hence it should be possible to introduce leptons without triggering proton decay and simultaneously keeping $SU(3)$ strong.

\subsection{Decoupling the flavor scale}
\label{sec:hyp}

A phenomenologically viable scenario must satisfy $\Lambda_{\rm F}\gg\Lambda$ in order to suppress ${\cal L}_{\rm SM,EFT}$. This is compatible with the generation of SM fermion masses if large RG enhancements come to our rescue. I therefore assume that $N_F$ can be chosen such that the $SU(3)$ dynamics in isolation has a strongly-coupled IR fixed point. According to a 2-loop analysis the conformal window is $9\leq N_F\leq16$, which suggests that for $N_F=9$ the fixed point is maximally strong (intriguingly, the same $N_F$  of my prototypical model). Lattice QCD methods are currently employed to establish the lower end, see e.g.~\cite{NF8} for a study of $N_F=8$ and for more references. 

In the conformal window, a negative anomalous dimension $\gamma$ for an operator in (\ref{EFT}) would enhance the corresponding Wilson coefficient at $\sim\Lambda$ by a factor $(\Lambda_{\rm CW}/\Lambda)^{|\gamma|}$, where  
\ba
\Lambda_{\rm CW}\sim \Lambda_{\rm F} e^{-2\pi/\beta_1\alpha(\Lambda_{\rm F})}, 
\ea
is the scale below which the theory becomes approximately conformal and $\beta_1=11-N_F*2/3$. For a maximally strong IR fixed point it is reasonable to expect $\gamma={\cal O}(1)$. Here I want to consider the most ambitious scenario, realized when some of the baryons of the strong $SU(3)$ theory have anomalous dimension $\gamma<-2$. There is no theoretical obstruction to this, since as I emphasized in the Introduction the only bound we can think of gives $\gamma>-3$. Needless to say, a lattice computation is required to prove this is truly realized in the present scenarios.

Under my assumption, (\ref{PC}) becomes {\emph{relevant}} below $\Lambda_{\rm CW}$ and we can safely take $\Lambda_{\rm F}$ as large as the Planck scale. Yet, because we typically expect $\lambda_Q(\Lambda_{\rm CW})\sim C_Q(\Lambda_{\rm CW}/\Lambda_{\rm F})^2\ll1$ I conclude that $\Lambda_{\rm CW}$ should be much larger than $\Lambda$ to allow sufficient RG evolution to generate the observed SM fermion masses (see Appendix~\ref{app:IRFP}). The UV parameters must therefore be chosen so that $\Lambda\ll\Lambda_{\rm CW}<\Lambda_{\rm F}\leq M_{\rm Pl}$.

The mass terms ${\cal L}_{\rm mass}$ are always a relevant deformation of the CFT and cannot be neglected at low energies. Similarly to~\cite{Luty:2008vs}, conformality is completely lost below 
\ba\label{lambda}
\Lambda\sim\Lambda_{\rm CW}\left({m_\psi}/{\Lambda_{\rm CW}}\right)^{1/(4-d)}, 
\ea
with $d$ the scaling dimension of the bilinear $\psi\overline{\psi}$ within the conformal window, and $m_\psi$ typically the largest constituent mass. %{\color{red} For definiteness I will assume that $m_\psi\sim m_{S_{2,3}}\gg m_{D,S_1,S'}$.} 
Note that $\Lambda$ can be naturally of order the TeV.

The operators in ${\cal L}_{\rm SM,EFT}, {\cal L}_{\rm gauge}$, and ${\cal L}_{\rm ETC}$ --- as well as those of higher engineering dimension --- remain irrelevant and play no role in my analysis. On the other hand, some of the self-interactions (\ref{self}) may be important. This is simultaneously dangerous and attractive. We certainly do not want interactions involving the light flavors $D,S,S'$ to become large in the IR, since one should then worry about their effect on the custodial symmetry, the Higgs mass, as well as the estimate of $\Lambda$ given in (\ref{lambda}). On the other hand, the possibility that some $\psi$-interactions become marginally relevant is a very welcome feature. In fact, these could introduce the desired ${\cal O}(1)$ violation of the family symmetry that is necessary to have flavor-dependent scaling dimensions $d_O$, and hence naturally account for the SM mass hierarchy (see the Introduction). Furthermore, they could drive the theory to a stronger fixed point with a small $d_O$ as in~\cite{Barnard:2013zea}. However, for simplicity I will assume that (\ref{self}) is not important, either because those operators remain irrelevant or because their coefficients are taken to be very small at the cutoff.

\section{Phenomenology}
\label{sec:pheno}

My formulation in terms of fundamental constituents is perturbative above $\Lambda_{\rm CW}$. The energy range $\Lambda<\mu<\Lambda_{\rm CW}$ may be qualitatively described by a (strongly-coupled) warped extra dimension thanks to the AdS/CFT correspondence. The effective 5D dual differs from the scenarios discussed in the literature, however. The bulk has a large gauge symmetry $\supset SU(N_F)^2\times U(1)_V$ broken on the IR to $SU(N_F)_V\times U(1)_V$ and on the UV by the gauging of the SM and tadpoles of adjoint bulk scalars (dual to the constituent mass operators). The SM baryon and lepton numbers, as well as $U(1)_{\rm T}$, are a subgroup of the bulk gauge symmetry, and are only broken at the boundaries by gauge anomalies. The SM fermions are embedded in {\emph{reducible}} representations, have large UV kinetic terms ($\lambda(\Lambda_{\rm CW})\ll1$), and are slightly peaked towards the IR ($\lambda$ is relevant). This still allows us to generate the SM mass hierarchy via wave-function overlap. Furthermore, the deconfined-confined transition in the early Universe is likely to be a smooth cross-over, which evades issues associated with a possible inflationary phase at temperatures below $\Lambda$ expected in warped 5D scenarios (see~\cite{Creminelli:2001th} and subsequent papers).

%\paragraph{Low energy theory}

The physics below the confinement scale $\Lambda$ is captured by a non-linear sigma model for the NGBs associated to the (approximate) symmetry breaking pattern $SU(N_F')\times SU(N_F')\to SU(N_F')_V$, where $N_F'<N_F$ is the number of light constituents at $\sim\Lambda$. To be concrete, we consider a scenario where 
\ba\label{s2s3}
m_\psi=m_{S_2}, m_{S_3}
\ea
drive confinement according to (\ref{lambda}). In this case there are $7$ light flavors ($T,D,S_1,S')$ and the $48$ pseudo-NGBs appear in the following representation of $SU(3)_c\times SU(2)_w\times SU(2)_{\rm cust}$:
\ba
{\bf 48}&=&{\bf (8,1,1)\oplus(3,2,1)\oplus(\overline{3},2,1)\oplus(3,1,2)\oplus(\overline{3},1,2)}\\\no
&\oplus&{\bf (1,3,1)\oplus(1,1,3)\oplus(1,2,2)\oplus(1,2,2)\oplus(1,1,1)\oplus(1,1,1)}.
\ea
Because of (\ref{aa}), baryon and lepton numbers are good symmetries, so none of these scalars couples to a lepton and quark pair, i.e. they are not composite lepto-quarks such as those studied in~\cite{Gripaios:2009dq}. 

All of the NGBs acquire a mass either from the gauge interactions or from (\ref{EFT}), which generically breaks all axial symmetries. The lightest states are typically the SM singlets, with an irreducible contribution to their masses squared of order $\sim m_{T,D,S,S'}\Lambda$. Decay rates into SM fermions are model-dependent. Yet, all of the exotic NGBs except the color triplets and the Higgs doublets can be singly produced and decay via couplings to the CP-odd SM gauge field operators $G\widetilde{G}, W\widetilde{W}, W\widetilde{B}, B\widetilde{B}$, that are unambiguously determined by the anomaly.

\subsection{Colored scalars}

TeV scale colored scalars are a generic prediction of QCD-like completions of PC and may well represent the first signature of Higgs compositeness. This is to be contrasted with most phenomenological approaches to PC, where color is assumed to be factorized in the global symmetry ${\cal G}$ of the strong dynamics thereby implying that this key signature is lost. An exception are warped GUT scenarios~\cite{Agashe:2004ci}\cite{Frigerio:2011zg}\cite{Barnard:2014tla}, where the entire SM is embedded into a simple group.~\footnote{$U(1)_{B,L}$ are broken when leptons and baryons reside in the same multiplet. This does not happen in the present models.} In the present paper unification is not required, though.

A characteristic feature of the QCD-like UV-completion we have been discussing is the presence of an accidental $U(1)_{\rm T}$ under which the exotic hadrons have integer charges (see table~\ref{t2}). The lightest state carrying ${\rm T}$-charge is collider-stable. It is reasonable to expect it is one of the color triplet NGBs ${\bf (3,2,1)}$ or ${\bf (3,1,2)}$:
\ba\label{TCh}
\overline{T}D,~~~~~~~~~~
\overline{T}S,~~~~~~~~~~\overline{T}S'.
\ea
(Baryons also have T-charge, but are typically heavier.) The lightest of (\ref{TCh}) is directly pair-produced or results --- accompanied by SM fermions --- from the decay of heavier composites. The latter processes are usually prompt, if decays into third generation SM fermions are kinematically allowed. The stable scalar then hadronizes into ${\rm T}$-hadrons by combining with one or two quarks. Because of (\ref{aa}) the ${\rm T}$-hadrons have fractional charges. 

${\rm T}$-hadrons are constrained by collider bounds on R-stops~\cite{Chatrchyan:2013oca}\cite{ATLAS:2014fka}. While the production at the LHC is similar, the larger energy deposition of ${\rm T}$-hadrons might push the current lower bound on the mass closer to the TeV. For small constituent masses, the T-hadron mass is of order $g_sf$, where $g_s$ is the QCD coupling at $\Lambda$, so we see that the LHC is already probing interesting regions of the parameter space. Constraints from searches of heavy nuclei and fractionally-charged matter can be very strong but less robust than collider bounds since they can be evaded either turning on higher dimensional operators that destabilize the exotic states on cosmological time scales, or simply postulating that reheating after inflation occurs at sufficiently low temperatures to significantly dilute their relic abundance. The destabilization might occur while preserving $U(1)_{\rm T}$, thus suggesting an intriguing link with the dark matter puzzle.

\subsection{Vacuum alignment}
\label{vacc}

Vacuum alignment represents a non-trivial issue due to the presence of exotic NGBs with SM charges. Fortunately, we can argue that color $SU(3)_c$ remains unbroken under very generic conditions~\cite{Vafa:1983tf}: the colored NGBs have positive mass squared of order $g_c^2f^2$ and trivial background values. The orientation of the vacuum is thus determined by the NGBs with electro-weak charges, and may be analyzed inspecting the physics of the symmetry breaking pattern 
\ba
SU(4)_L\times SU(4)_R\to SU(4)_V.
\ea
The corresponding NGBs decompose under $SU(2)_w\times SU(2)_{\rm cust}$ as ${\bf 15}=({\bf 2},{\bf 2})+({\bf 2},{\bf 2})+({\bf 3},{\bf 1})+({\bf 1},{\bf 3})+({\bf 1},{\bf 1})$, that is two doublets ($H_i$), a weak triplet ($\phi$), a custodial triplet ($\phi'$), and a singlet ($\eta$). We introduce the NGB matrix $U=e^{i\sqrt{2}\Pi/f}\to LUR^\dagger$ ($L\in SU(4)_L$, $R\in SU(4)_R$)
\ba\label{NGB}
\Pi=\left( \begin{array}{cc}  \phi_a\sigma_a+\frac{1}{\sqrt{2}}\eta {\bf 1} & {\cal H}_1+i{\cal H}_2 \\ {\cal H}_1^\dagger-i{\cal H}_2^\dagger & \phi'_a\sigma_a-\frac{1}{\sqrt{2}}\eta {\bf 1} \end{array}\right),
\ea
where %${\cal H}_i\equiv h_i^\alpha\sigma^\alpha$, with $\sigma^\alpha=(i{\sigma}^a,1)$. Note that 
${\cal H}_i=\widetilde{H_i} H_i$ and $H_i$ have hyper-charge ${-1/2}$. The kinetic term is $\frac{f^2}{4}{\rm tr}(D_\mu U^\dagger D^\mu U)$. We can combine ordinary charge conjugation and parity to an $SU(4)_V$ rotation and define a CP so that $H_1$ is even.~\footnote{We can choose parity as usual, $U(x)\to U^\dagger(x_P)$, whereas charge conjugation as $U(x)\to CU^t(x) C^\dagger$, where $C={\rm diag}\left(-\epsilon, \epsilon\right)$. The various NGB components transform as $(\phi^{(')},\eta, h_1,h_2)\to(-\phi^{(')},-\eta,-h_1,-h_2)$ under parity, and as $(\phi^{(')},\eta, h_1,h_2)\to(-\phi^{(')},\eta,- h_1,+h_2)$ under C. Combining the two we see that the fields $\phi,\phi',h_2,\eta$ are odd whereas $h_1$ is even. The couplings to the SM quarks generically violate all discrete transformations explicitly. 
}

The gauge interactions will favor $H_i=\phi=\phi'_\pm=0$, whereas $\eta, \phi'_0$ remain flat directions. Vacuum alignment is thus controlled by the PC couplings~(\ref{PC}), and specifically by those associated to the top quark. In what follows we turn to a discussion of the latter.

%We will now show that the custodial symmetry is generically broken by the vev of the exotic electro-weak doublet and triplets. Yet, there exist natural values of the parameters $C_q,C_u'$ that effectively turn the model into a custodial-symmetric $SU(4)/Sp(4)$ scenario. It is then possible to argue that $v^2<f^2$, a $126$ GeV Higgs, and a small electro-weak $T$ parameter can be realistically obtained.

\subsubsection{A choice of top partners}

The top partners $O$s are generally in reducible representations of the chiral symmetry. The spurious $SU(4)_L\times SU(4)_R$ charges of the mixings $\lambda$ depend on which of the explicit expressions in table~\ref{t} dominate, and can be $\lambda\sim({\bf1},{\bf 4}),~({\bf1},{\bf 6}+{\bf10}),~({\bf6}+{\bf10},{\bf1}),~({\bf4},{\bf4})$.

Ultimately, the choice of representation depends on unknown UV physics and is model-dependent. We found that in the bulk of the parameter space the potential of the NGBs arises at ${\cal O}(\lambda^2)$, which is somewhat disfavored by considerations on the naturalness of the weak scale,~\footnote{In the case $\lambda$ is charged under both $SU(4)_{L,R}$ the potential arises at $\lambda^2$. On the other hand, if $\lambda$ is charged only under either $SU(4)_L$ or $SU(4)_R$, no potential is induced.} and furthermore in many models the custodial $SU(2)_w\times SU(2)_{\rm cust}\subset SU(4)_V$ may be spontaneously broken, leading to an unsatisfactory phenomenology.~\footnote{Some PC couplings lead to an $SU(2)_{\rm cust}$ violating mixing between the two Higgses ${\rm tr}\left({\cal H}\sigma_3{\cal H}^\dagger\right)=2iH_2^\dagger H_1+{\rm hc}$ (${\cal H}\equiv{\cal H}_1+i{\cal H}_2$) that triggers a custodial-breaking vacuum $v_2\sim v_1\sim v$ which, from the non-linearities of the sigma model, results in $\widehat T\sim v^2/f^2$~\cite{Gripaios:2009pe}\cite{Mrazek:2011iu}.} Rather than performing a general analysis I therefore decided to focus on a specific scenario with some interesting physics. 

%The important point to us is that there exist scenarios with a realistic phenomenology. 

I will now argue that under technically natural assumptions about the UV we can ``choose" a set of top-partners that lead to a realistic phenomenology with a light Higgs and small electro-weak $T$ parameter. Our choice of the $t_L$-partner corresponds to $\lambda_q\sim({\bf{1},\bf{6}})$ and/or $({\bf{6},\bf{1}})\in SU(4)_L\times SU(4)_R$:
\ba\label{hypq1}
O_q=(\overline{T\psi})\overline{\psi}_\alpha~~{\rm and/or}~~(\psi\psi)^\dagger\overline{T}_\alpha
\ea
with $\psi=D,S$. We then choose $O_u$ oriented dominantly along one of the two $SU(2)_w\times SU(2)_{\rm cust}$-singlet combinations in $(\overline{T\psi})^\dagger\psi\sim({\bf 4},{\bf 4})$ (again $\psi=D,S$), up to corrections proportional to the small parameter $\delta$: 
\ba\label{hypu1}
O_u=w_1(\overline{TD})^\dagger D+w_2(\overline{TS})^\dagger S'-w_3(\overline{TS'})^\dagger S,~~~~~~~~~w_{1,2,3}=1+{\cal O}\left(\delta\right).
\ea
Equivalently, we can view $\lambda_{q,u}$ as spurions transforming under $SU(4)_L\times SU(4)_R$ as $\lambda_q\to L\lambda_qL^t$ and/or $\lambda_q\to R\lambda_qR^t$ and having background:
\ba\label{hypq}
(\lambda_{q})_i\equiv\widetilde\lambda_q^{(A)}\left( \begin{array}{cc}  
{\bf0}~ & Q_i \\
-Q_i^t~ & {\bf0}
\end{array}\right)+
\widetilde\lambda_q^{(S)}\left( \begin{array}{cc}  
{\bf0}~ & Q_i \\
Q_i^t~ & {\bf0}
\end{array}\right),
\ea
where $i=1,2$ is the $SU(2)$ index carried by $q_i$ and $Q_1=\frac{1}{2}(\sigma_1+i\sigma_2)$, $Q_2=\frac{1}{2}({\bf1}-\sigma_3)$, and $\lambda_u\to R^*\lambda_uL^\dagger$ with
\ba\label{tn}
\lambda_u\equiv\widetilde\lambda_u\Upsilon\left[1+{\cal O}(\delta)\right],~~~~~~~~~~~\Upsilon\equiv \left( \begin{array}{cc}  
\epsilon~ & {\bf0} \\
{\bf0}~ & \epsilon
\end{array}\right),
\ea
%Also, $\widetilde\lambda_q^{(A),(S)}$ are aligned in SM family space. 
The choice (\ref{hypq}) is a legitimate assumption about the UV dynamics because equivalent to postulating that the operators of the type $\overline{D}(TS)^\dagger$ and $\overline{S}(TD)^\dagger$ in table \ref{t} have suppressed coefficients. A careful look reveals that the same is true for $\lambda_u$. Indeed, (\ref{tn}) is technically natural because the coupling $\lambda_u{u}O_u$ with $\delta=0$ respects an $SU(4)_*$ defined by $({\bf 4},{\bf1})={\bf 4}$, $({\bf1},{\bf 4})={\bf \bar 4}$, and acting as $L=V$, $R= \Upsilon V^* \Upsilon^\dagger$, $V\in SU(4)_*$. Note that $SU(4)_*\subset SU(4)_L\times SU(4)_R$ resides partly in the vector and partly in the axial components of the chiral symmetry. More precisely, $\lambda_u$ is invariant under an $Sp(4)_V\in SU(4)_V$ when $\delta=0$; the remaining five generators of $SU(4)_*$ are in the axial part of $SU(4)_L\times SU(4)_R$ and coincide with the generators associated to $H_1$ and $\eta$.

What we find is quite amusing. First, the choice $\delta\ll1$ in (\ref{tn}) is motivated by some UV symmetry $SU(4)_*$. Second, the low energy theory is described by the reduced coset $SU(4)_*/Sp(4)_V$, with $H_1,\eta$ exact NGBs, and has a light Higgs and a small $T$-parameter. To appreciated this latter point observe that, neglecting ${\cal O}(\lambda^4)$ contributions to the potential (to be studied in the next subsection), our choice (\ref{hypq1}) (\ref{hypu1}) generates the following NGB potential:
\ba\label{Vu}
\delta V&=&C_u~{\rm tr}\left[(\lambda_uU)(\lambda_uU)^*\right]\\\no
&\propto&|\widetilde\lambda_u|^2\left\{\phi_a^2+\phi_a'^2+2h_2^2+{\cal O}(\delta,\Pi^3)\right\},
\ea
where $C_u$ is positive (see Appendix~\ref{app:C}). Hence, the mixing $\lambda_u$ results in a {\emph{positive}} mass squared for $\phi, \phi',h_2$ of order $m^2_{\rm heavy}\sim\widetilde\lambda^2_u f^2$, whereas $h_1,\eta$ remain exactly massless as long as $\delta=\lambda_q=0$. The potential $\delta V$ is minimized at $U=1$, and hence respects the $Sp(4)_V\subset SU(4)_*$. Below $m_{\rm heavy}$ the relevant symmetry-breaking patter is $SU(4)_*\to Sp(4)_V$, as anticipated, and the dynamics is described by a non-linear sigma model for $H_1, \eta$, parametrized by $\Sigma\equiv U\Upsilon\to V\Sigma V^t$. With (\ref{hypq}) (\ref{tn}) and $\delta\ll1$ the corrections to the electro-weak $T$ parameter can be naturally small because induced by controllably small parameters $\delta,\lambda_q/4\pi$.

\subsubsection{The Higgs potential}
\label{sec:V}

Because $\phi,\phi',H_2$ have large positive masses, for small $\delta,\lambda_q$ the only NGBs that can potentially get a sizable vacuum are the SM Higgs boson $H_1$ and $\eta$. The singlet obtains a potential from subleading $\lambda$ couplings and $m_{D, S, S'}\neq0$. We assume that these effects are such that its vacuum is trivial. We therefore set $\eta=0$ and take $H_1=(0,h/\sqrt{2})^t$, so the NGB matrix simplifies into 
\ba
U=\left( \begin{array}{cc}  
{\bf1}c_h~ & i{\bf1}s_h \\
i{\bf1}s_h ~& {\bf1}c_h
\end{array}\right),
\ea
where we defined $s_h=\sin(h/f)$ and $c_h=\cos(h/f)$, whereas ${\bf1}$ is the 2 by 2 identity. The 2-derivative non-linear sigma model gives $m_W^2(h)=\frac{1}{4}g^2f^2s_h^2=\cos^2\theta_w m_Z^2(h)$ and hence $\xi\equiv{v^2}/{f^2}=\langle s_h^2\rangle$. Furthermore, for our choice of $O_{q,u}$, see (\ref{hypq1}) (\ref{hypu1}), one finds (from ${\rm tr}[U^*\lambda_u\lambda_q]$ when $\lambda_q\to L\lambda_q L^t$ and ${\rm tr}[U\lambda_q\lambda_u]$ when $\lambda_q\to R\lambda_q R^t$) $m_t(h)={y_t}fs_h/{\sqrt{2}}$ with $y_t\sim\lambda_q\lambda_u{\sqrt{N_c}}/{4\pi}$. The Higgs couplings to the massive vector bosons and to the top quark deviate from the SM by a factor $\sqrt{1-\xi}$~\cite{Giudice:2007fh}.

In our model the Higgs potential is controlled by $\delta, \lambda_q/4\pi$ and schematically reads %To obtain it we first integrate out all fluctuations (including the SM) down to momenta of order $\Lambda$. Then we evolve the potential down to scales of order $\sim m_t$, where we match onto the SM. Summing both UV and IR contributions, the Higgs potential becomes:
\ba\label{Vh}
V_h=\alpha s_h^2+\beta s_h^4+\gamma s_h^4\ln s_h^2+{\cal O}(s_h^6).
\ea
Our expansion in $s_h^2\ll1$ is justified because we are interested in vacua with $0<s_h^2\ll1$. In that limit, besides the trivial vacuum, we find:
\ba\label{solu}
\xi\equiv\langle s_h^2\rangle=-\frac{\alpha}{2\beta+\gamma+2\gamma\ln\xi},~~~~~~~~~~~~m_h^2=\frac{\xi(1-\xi)}{f^2}\left[8\beta+12\gamma+8\gamma\ln \xi\right].
\ea
When $\alpha<0$ the latter solution is always favored compared to $s_h=0$. Issues of global stability (in particular with respect to a possible extrema at $s^2_h=1$) depend on the incalculable coefficient of higher powers in $s_h^2$. For simplicity, in what follows we will assume the latter are such that $s_h=1$ is disfavored compared to (\ref{solu}).

The log in (\ref{Vh}) arises from loops of the top quark, and is numerically negligible compared to $\beta$ unless $\xi$ is unnaturally small. I will neglect it for simplicity. The Higgs potential receives important UV contributions from fluctuations above $\Lambda$. The results of the previous section imply that polynomials of $\lambda_u$ alone can only contribute proportionally to powers of the small parameters $\delta$. Moreover, since $\lambda_q$ is charged under either $SU(4)_{L}$ or $SU(4)_{R}$, it follows that $\lambda_q$ will necessarily appear accompanied by $\lambda_u$. Up to $O(\lambda^4)$ we find:~\footnote{A perturbative series in $\lambda$ is meaningful if $\lambda\ll4\pi/\sqrt{3}$. Given that $y_t\sim\lambda_q\lambda_u/4\pi$ and assuming $\lambda_q\sim\lambda_u$, the largest possible value for $\lambda_{q,u}$ is of order $\sqrt{4\pi}$, which we believe still allows a reasonable perturbative expansion.} 
\ba\label{alpha}
\alpha=\hat\alpha \Lambda^2f^2\frac{N_cy_t^2}{16\pi^2},~~~~~~~~~~~\hat\alpha=\hat\alpha_1\left(\frac{4\pi\delta}{\sqrt{N_c}\widetilde\lambda_u}\right)^2+\hat\alpha_2\left(\frac{\widetilde\lambda_q}{\widetilde\lambda_u}\right)^2+\cdots,
\ea
where $\hat\alpha_i$ are numbers of order unity, the top Yukawa is renormalized at $\Lambda$, and the dots refer to subheading terms. To obtain $\xi\ll1$, the parameter $\hat\alpha$ must be somehow tuned to a value smaller than unity. Fortunately, in the present model a cancellation is possible provided $|\delta|\sim\frac{\sqrt{N_c}}{4\pi}|\widetilde\lambda_{q}|$. As shown above, this is a technically natural assumption. What is unnatural here is that $\delta,\lambda_q$ conspire so that $\hat\alpha$ is much smaller than unity: this is the usual fine-tuning problem characterizing these scenarios.~\footnote{Because the Higgs quartic is naturally of the right order (see below), this is the only tuning in the Higgs potential. A posteriori we estimate it to be of order ${\Lambda^2}/{m^2_t}$, with $\Lambda$ the mass of the fermionic resonances interpolating the top partners $O_{q,u}$. The latter should be light to reduce the tuning. On the other hand, as I will show below, $\Lambda\sim4\pi f$ can still be compatible with a $m_h=126$ GeV Higgs.}

We found a unique contribution to $\beta$ at ${\cal O}(\lambda^4)$ from the $\lambda_u^4$ term, which is consistently proportional to $\delta^2$. Because $|\delta|\sim\frac{\sqrt{N_c}}{4\pi}|\widetilde\lambda_{q}|$, this is secretly ${\cal O}(y_t^4)$ and comparable to the top-quark IR effect. Other, larger contributions to $\beta$ may arise either at $\lambda^6$ or $\lambda^8$. Taking $\lambda^2_q\sim\lambda^2_u\sim4\pi/\sqrt{N_c}$ for definiteness, this means
\ba\label{b}
\beta=\hat\beta\Lambda^2f^2\frac{N_cy_t^2}{16\pi^2},~~~~~~~~~~\hat\beta=\hat\beta_1 \frac{\sqrt{N_c}y_t}{4\pi}+\hat\beta_2 \left(\frac{N_cy_t^2}{16\pi^2}\right)+\cdots,
\ea
where $\hat\beta_i$ are numbers naturally of order one. Plugging into (\ref{solu}) and using $\Lambda=4\pi/\sqrt{N_c}$ we see that the physical Higgs mass can be naturally at $m_h=126$ GeV for $\hat\beta_i={\cal O}(1)$. Two prototypical examples are $\hat\beta_1\sim0.5, \hat\beta_2=0$ and $\hat\beta_1=0, \hat\beta_2\sim4$.

\subsection{Flavor violation}

Flavor violation in this model is analogous to that discussed in previous phenomenological studies of PC. In particular, important sources of flavor-violation beyond the SM are generated at $\sim\Lambda$ by (\ref{PC}). These have been shown to be under control for $\Lambda\gtrsim10$ TeV (see e.g.~\cite{KerenZur:2012fr}). However, our model presents two important differences compared to the scenarios discussed in the literature. 

First, in generic realizations the ``flavorful" scalars $\overline{D}S_I$ couple to quark bilinears and generate FCNC 4-fermion operators at tree-level. These processes may be parametrically enhanced compared to those present in standard scenarios of PC when the flavorful scalars are lighter than $\Lambda$. However, under our assumption that (\ref{s2s3}) induce confinement via (\ref{lambda}), $\overline{D}S_{2,3}$ acquire masses of order $\Lambda$ ($\overline{D}S_1$ is the Higgs doublet), and therefore this potential problem is evaded.

Second, the SM fermions generically couple to more than one PC operator in (\ref{PC}). This may lead to unacceptably large deviations in precision flavor observables. To robustly avoid a proliferation of flavor-violating parameters we may invoke flavor symmetries. Alternatively, we can assume that the UV dynamics is such that a single structure in each column of table~\ref{t} has unsuppressed coefficients. For example, we assume that the dominant flavor-violating parameters in (\ref{PC}) are $(C_q)_{iI}, (C'_u)_{iI}, (C_d)_{iIJ}$, where  $i=1,2,3~(I,J=1,2,3)$ are the $q,u,d$ ($S$) family indices. If these Wilson coefficients have a hierarchical structure at $\Lambda$ similar to that predicted in warped 5D scenarios, the constraint $\Lambda\gtrsim10$ TeV should be enough to satisfy current bounds. It is evident that there are regions of the parameter space where flavor violation beyond the SM is under control.

\section{Outlook}
\label{sec:conclusions}

I presented a description of Partial Compositeness (PC) in terms of an $SU(3)$ gauge theory with $N_F$ Dirac flavors in the fundamental representation. These models feature:
\begin{itemize}
\item fermionic partners for all SM representations ($S'$ is superfluous in this respect, but required to have a custodial symmetry);
\item automatic suppression of proton decay (see (\ref{aa}));
\item absence of Landau poles below the Planck scale;
\item a strongly-coupled IR fixed point, naturally broken at $\Lambda\ll\Lambda_{\rm CW}$ by the constituent masses (a walking behavior might also be a viable option);
\item a light NGB Higgs, a custodial symmetry, and a realistic vacuum alignment;
\item distinctive collider signatures, including stable T-hadrons with fractional charges.
\end{itemize}
The Lagrangian (\ref{EFT}) may be generated by the exchange of heavy particles at a flavor scale $\Lambda_{\rm F}$. Therefore this picture should be seen as an effective field theory below $\Lambda_{\rm F}$. 

The point of this paper is that such an EFT may actually become a fully fledged UV-completion of the SM under very reasonable conditions on the non-perturbative dynamics. Indeed, when the scaling dimension $d_O$ of some $SU(3)$ baryons is smaller than $2.5$ in the conformal window, then the linear couplings (\ref{PC}) to the SM fermions become marginally relevant, that is the PC interactions (\ref{PC}) are {\emph{dangerous irrelevant}} operators that cannot be ignored in the IR. If this key condition is met it becomes possible to take $\Lambda_{\rm F}$ as large as the Planck scale and still obtain realistic Yukawa couplings for the NGB Higgs. In this limit {\emph{all}} SM fermions must acquire a mass from the PC interactions (\ref{PC}), which in our model is possible with $N_F\geq9$. Importantly, we have argued that under reasonable assumptions on the Wilson coefficients in (\ref{PC}) all phenomenological constraints are under control for $f\gtrsim1-2$ TeV.

Lattice QCD techniques should be used to verify whether an anomalous dimension of order $-2$ is possible for at least one of the two sets of operators in table~\ref{t}. The results of Appendix \ref{app:IRFP} suggest that in realistic models (\ref{PCi}) is a small perturbation of the CFT even for $d_O<2.5$, implying that the dynamics we have to simulate on the lattice is simply a familiar $SU(3)$ gauge theory with $N_F$ light flavors, up to the weak gauging of the SM and small ${\cal O}(\lambda^2)$ corrections.~\footnote{The most promising way to probe the hypothesis $d_O\lesssim2.5$ is to directly extract $d_O$ from the correlators of $O$. Such a study is complicated by the limited range of validity of the lattice, which unfortunately precludes an exploration of the exactly conformal regime. Inferring $d_O$ from a study of the spectrum in a theory with (\ref{PCi}) turned on, similarly to what discussed in~\cite{Luty:2008vs}\cite{Sannino:2008nv}\cite{DeGrand:2009mt}\cite{DelDebbio:2010jy} for non-abelian theories with massive fermions, does not seem to be very convenient because for any $d_O\lesssim2.5$ and $d_O>2.5$ the IR dynamics is corrected by tiny effects ${\cal O}(\lambda^2)$ with a moderate dependence on the scaling dimension. To extract $d_O$ from the spectrum, (\ref{PCi}) should be so strongly relevant to invalidate (\ref{beta}).}

Asymptotically-free non-abelian realizations of PC can still be phenomenologically relevant in less ambitious realizations, though. While $d_O<2.5$ is certainly ideal, it is important to realize that $d_O\sim2.5$ is sufficient to let $\Lambda_{\rm F}\gg\Lambda$. In fact, analogously to what argued in~\cite{Luty:2004ye} for flavor scenarios of the type $QQ\Phi$, the scale at which the top quark couplings $\lambda_{L, R}$ become of order $Y_*\sim g_*\sim4\pi$ can be estimated from (\ref{yt}) as 
\ba
\Lambda\left(\frac{4\pi}{y_t}\right)^{1/(d_{O_L}+d_{O_R}-5)}.
\ea
For $d_{O_L}=d_{O_R}=2.6-2.7$ this is a factor $10^3-10^6$ larger than $\Lambda$, which could be enough to suppress the most dangerous FCNC effects from (\ref{ETC}) and ${\cal L}_{\rm SM,EFT}$.  As opposed to the setup with larger $\Lambda_{\rm F}$, in these cases one may also consider models where {\emph{only}} the top Yukawa (and perhaps the bottom as well) is induced by PC, whereas the light SM fermions arise from (\ref{ETC}).

%Interestingly, QCD-like completions of PC lead to a distinctive phenomenological prediction: ${\rm T}$-hadrons. We have seen that the hyper-charge quantum numbers of the fields in table \ref{m1} can be chosen such that the SM baryon and lepton numbers arise as accidental global symmetries, as in the SM. With these charge assignments the lightest NGB color-triplet is always collider-stable. After being pair-produced, the scalar confines into exotic long-lived ${\rm T}$-hadrons with fractional charges, leading to novel signatures of TeV scale compositeness. 

The class of theories discussed in this paper is only one of the possible UV completions of PC. There exist other interesting directions to explore. These include the non-abelian models of~\cite{Ferretti:2013kya}, scenarios with relevant 4-$\psi$ interactions~\cite{Barnard:2013zea} or with strongly-coupled scalars~\cite{Strassler:2003ht}. Alternatively, one could relax the assumption that the $Q$s are weakly-coupled and envision a situation where, say, $SU(3)_c$ is embedded into a stronger group: in this framework the anomalous dimension of $QO$ receives an additional negative contribution that makes $\lambda$ more relevant. It is a little premature to determine which of these approaches offers the most compelling realization of PC. For this reason lattice simulations should ideally be performed on as many theories as possible.

\acknowledgments

I would like to thank the organizers of the 2015 workshop ``Lattice for Beyond the Standard Model Physics" at LLNL, where this work was first presented. I am indebted with the participants for interesting conversations and for encouraging me to publish my work. I am especially grateful to K. Agashe for sharing with me his perspective on various topics related to the NGB Higgs and PC as well as for comments on the manuscript. I also thank G. Ferretti and D.~Karateev for correspondence. This work was supported in part by NSF Grant No. PHY-1315155 and by the Maryland Center for Fundamental Physics.

\appendix

\section{RG evolution of $\lambda$}
\label{app:IRFP}

Let us work in Euclidean space, and consider the following CFT deformation
\ba\label{Sint}
S_{\rm int}=\int d^4x~\lambda\overline{Q}O+{\rm hc},
\ea
with $Q$ a SM fermion and $O$ a CFT operator of dimension $d_O=2.5+\gamma_O$. We are interested in the running of the coupling $\lambda$. I will work in the regime ${g^2}/{16\pi^2}\ll|\gamma_O|$, that allows us to neglect the effect of the SM gauge interactions. Under this hypothesis there is no vertex correction for (\ref{Sint}). Moreover, since divergences are local, the 2-point function of $O$ is not renormalized. However, the wave-function of $Q$ receives a divergent contribution at ${\cal O}(\lambda^2)$ from $\langle O\overline{O}\rangle$. Indeed, the operator product expansion gives:
\ba\label{2point}
O(x)\overline{O}(y)\to \frac{C}{2\pi^2}\frac{(x-y)_\mu\Gamma^\mu}{|x-y|^{2d_O+1}}+\cdots,
\ea
with $C>0$ by unitarity, $\left\{\Gamma^\mu,\Gamma^\nu\right\}=2\delta^{\mu\nu}$ and the dots referring to less singular terms. The Fourier transform of (\ref{2point}) is divergent in the limit $\gamma_O\to0$. The cutoff-dependence is removed by wave-function renormalization, $Q=\sqrt{Z_Q}Q_{\rm r}$. The renormalized coupling is finally $\lambda_{\rm r}=\sqrt{Z_Q}\lambda\mu^{\gamma_O}$, where $\mu$ is an arbitrary IR renormalization scale, and the beta function reads  
\ba\label{beta}
\frac{d\lambda_{\rm r}}{d\ln \mu}&=&\gamma_O\lambda_{\rm r}+\gamma_Q\lambda_{\rm r}+{\cal O}(g^2\lambda_{\rm r}).
\ea
To proceed we first calculate $\gamma_Q$ in the limit $\lambda_{\rm r}\ll1$ and $|\gamma_O|\ll1$. I will then comment on a generalization of the result.

The wave-function $Z_Q$ may be found by inspecting the divergent part in the path integral. At quadratic order in the perturbation I find:
\ba
e^{-S_{\rm int}}&\supset&+\frac{1}{2}\lambda_{\rm r}^2\mu^{-2\gamma_O}\int d^4x\int d^4y~\overline{Q_{\rm r}}(x)O(x)\overline{O}(y)Q_{\rm r}(y)+\cdots\\\no
%&\mapsto&+\frac{C}{2\pi^2}\frac{1}{2}\lambda_{\rm r}^2\mu^{-2\gamma_O}\int d^4x\int d^4w~\overline{Q_{\rm r}}(x)\frac{(-w_\mu)\Gamma^\mu}{|w|^{2d_O+1}}Q_{\rm r}(x+w)+\cdots\\\no
&\mapsto&-\frac{C}{2\pi^2}\frac{1}{8}\lambda_{\rm r}^2\mu^{-2\gamma_O}\int d^4w~\frac{1}{|w|^{2d_O-1}}\int d^4x~\overline{Q_{\rm r}}(x)\slashed{D}Q_{\rm r}(x)+\cdots
\ea
where after $\mapsto$ I retained only the most singular piece using (\ref{2point}) with $y=x+w$, then Taylor-expanded $Q_{\rm r}(x+w)$ at first order in $w$; the zeroth order leads to a vanishing integral while higher orders are regular as $|w|\to0$. For $\gamma_O\to0$ the integral in $w$ is UV divergent and the cutoff-dependence can be reabsorbed into $\delta Z_Q=Z_Q-1$, which at the order we are working appears in $e^{-S_{\rm int}}$ as $-\delta Z_Q\int d^4x~\overline{Q_{\rm r}}(x)\slashed{D}Q_{\rm r}(x)$. It is readily seen that
\ba
\delta Z_Q^{\rm div}=-\frac{C}{16\pi^2}\lambda_{\rm r}^2\mu^{-2\gamma_O}\int_{1/\Lambda_{\rm UV}}^{1/\mu} d^4w~\frac{1}{|w|^{2d_O-1}}=\frac{C}{8}\lambda_{\rm r}^2\ln \frac{\mu}{\Lambda_{\rm UV}}\left(1+{\cal O}(\gamma_O)\right),
\ea
plus corrections of order $\lambda_{\rm r}^4, g^2$. Hence:
\ba\label{and}
\gamma_Q=\frac{C}{16}\lambda_{\rm r}^2+{\cal O}(\lambda_{\rm r}^4, \lambda_{\rm r}^2\gamma_O,g^2).
\ea
Plugging this result into eq.~(\ref{beta}) we get:
\ba\label{RG}
\lambda_{\rm r}(\mu)=\frac{\lambda_0\left(\frac{\mu}{\mu_0}\right)^{\gamma_O}}{\sqrt{1-\frac{C}{16}\frac{\lambda^2_0}{\gamma_O}\left(\frac{\mu}{\mu_0}\right)^{2\gamma_O}}}.
\ea
with $\mu_0<\Lambda_{\rm CW}$ a reference scale. For $\gamma_0<0$ the RG evolution admits a non-trivial IR fixed point $\lambda_{\rm r}^2=-\frac{16}{C}{\gamma_O}$. If the IR fixed point is not reached, or if $\gamma_O>0$, then the second term in (\ref{beta}) is negligible and $\lambda_{\rm r}(\Lambda)\simeq\lambda_0\left({\Lambda}/{\mu_0}\right)^{\gamma_O}$. As a result, the SM Yukawa (\ref{yt}) may naturally manifest a hierarchical structure whenever the $\gamma_O$s are flavor-dependent, suggesting a possible origin for the SM flavor structure.

In models with $\lambda_{\rm r}(\mu_0)\ll1$ the largest value for (\ref{yt}) is obtained when the couplings $\lambda_{L,R}$ reach the fixed point. For instance, normalizing the operators $O$ as in the Introduction gives $C\sim1/g_*^2$ and $y_t\sim Y_*\gamma_O$ (for simplicity I assumed a comparable scaling dimension for the partners of both $t_{L,R}$). In a strong dynamics with $Y_*\sim4\pi$ the top Yukawa can thus be obtained with a $|\gamma_{O}|$ small enough to trust (\ref{RG}). 

Yet, my analysis generalizes to larger $|\gamma_O|$ if $O$ has suppressed 4-point functions. In the latter case the ${\cal O}(\lambda_{\rm r}^4)$ terms in (\ref{and}) are subleading and $\gamma_Q\propto\lambda_{\rm r}^2$, again with a positive coefficient by unitarity. The same RG evolution as in (\ref{RG}) is obtained despite $\lambda_{\rm r}$ may now grow as large as $g_*,Y_*$. This is the case realized in warped 5D models, see e.g.~\cite{Contino:2003ve}\cite{Contino:2004vy}\cite{Agashe:2004rs}.

While the running of $\lambda$ can be understood in generality, it is not possible to precisely assess the impact of $\lambda$ on the couplings of the CFT. It is natural to expect that (\ref{Sint}) will violate the CFT introducing corrections of order $\lambda^2/16\pi^2$ on all beta functions, the largest effect being obtained when $\lambda$ is relevant. For example, in a non-abelian theory with coupling $\alpha_{\rm CFT}=g^2_{\rm CFT}/4\pi$ I estimate a correction 
\ba\label{pertL}
\frac{\Delta\alpha_{\rm CFT}}{\alpha_{\rm CFT}}\leq \kappa\frac{g_*^2}{16\pi^2}\frac{\alpha_{\rm CFT}}{4\pi}|\gamma_O|\ln\mu/\Lambda_{\rm FP}
\ea
where $\Lambda_{\rm FP}>\mu$ is the scale at which $\lambda$ approaches the fixed point and $\kappa={\cal O}(1)$. The effect should be relatively small for the $|\gamma_O|$s we are interested in.

\section{Positivity of the coefficient $C_u$}
\label{app:C}

Let us consider ${\cal L}\supset\overline{u_R}\lambda_uO+{\rm hc}$, with $O$ transforming as a ${\bf 4}\times{\bf 4}={\bf 6}+{\bf 10}$. Note that for convenience in this Appendix we use a Dirac fermion notation.

Within the effective field theory, the most general flavor-invariant Lagrangian quadratic in $\lambda_u$ reads
\ba
{\cal L}&\supset&C_1~{\rm tr}\left[(\lambda_uU)(\lambda_uU)^\dagger\right]+C_2~{\rm tr}\left[(\lambda_uU)(\lambda_uU)^*\right].
\ea
This is the potential discussed at the end of Section \ref{vacc}. We can find an explicit expression for $C_{1,2}$ by matching the second derivative of the partition function (in Minkowski space) in the EFT and in the fundamental theory. We work at all orders in the strong dynamics in the $SU(4)_V$-symmetric vacuum $U=1$. We neglect the SM gauge interactions and expand up to $O(\lambda_u^2)$. 

Within the EFT we find:
\ba\label{ZEFT}
\left.\frac{\partial^2\ln {\cal Z}}{\partial (\lambda_u)_{ij}\partial (\lambda^*_u)_{kl}}\right|^{\rm EFT}_{\lambda_u=0}&=&i\int d^4y~\left[C_1\delta_{ik}\delta_{jl}+C_2\delta_{il}\delta_{jk}\right].
\ea
In the fundamental theory:
\ba\label{A}
\left.\frac{\partial^2\ln {\cal Z}}{\partial (\lambda_u)_{ij}\partial (\lambda^*_u)_{kl}}\right|_{\lambda_u=0}&=&-\int d^4x\int d^4y~\langle\overline{u_R(x)}O_{ij}(x)\overline{O_{kl}}(y)u_R(y)\rangle\\\no
&=&+\int d^4x\int d^4y~{\rm tr}\left[\langle u_R(y)\overline{u_R(x)}\rangle\langle O_{ij}(x)\overline{O_{kl}}(y)\rangle\right].
\ea
The flavor indices $i,j,k,l$ are in the fundamental of $SU(4)_V$. In the second step we used the fact that the fields anti-commute (the trace acts on the spinor indices) and that under our simplifying assumptions the correlators factorize.

We now use a spectral decomposition for the 2-point function:
\ba
\langle O_{ij}(x)\overline{O_{kl}}(y)\rangle&=&\left(\delta_{ik}\delta_{jl}-\delta_{il}\delta_{kj}\right)\int_0^\infty ds~\rho(s)\int\frac{d^4p}{(2\pi)^4}e^{ip(x-y)}i\frac{\slashed{p}+\sqrt{s}}{p^2-s}\\\no
&+&{\rm symmetric},
\ea
where $\rho(s)\geq0$. A completely analogous expression holds for $u_R$ provided we take $\rho(s)=\delta(s)$ and replace the round parenthesis with 1. Substituting these expressions in (\ref{A}) and performing the integrals we arrive at
\ba\label{ZUV}
\left.\frac{\partial^2\ln {\cal Z}}{\partial (\lambda_u)_{ij}\partial (\lambda^*_u)_{kl}}\right|_{\lambda_u=0}&=&i\int d^4y~C_u\left[\delta_{ik}\delta_{jl}-\delta_{il}\delta_{kj}\right]+{\rm symmetric},
\ea
with
\ba
C_u=4\int\frac{d^4p_E}{(2\pi)^4}\int ds\frac{\rho(s)}{p_E^2+s}>0.
\ea
The symmetric part (i.e. the ${\bf 10}\in SU(4)_V$ component) will not be important to us because we are interested in the limit $\delta=0$ of (\ref{tn}) (i.e. $\lambda_u$ is anti-symmetric).

Finally, matching (\ref{ZEFT}) and (\ref{ZUV}) implies $C_1=-C_2=C_u>0$. %This result allows us to conclude that the NGBs $\phi,\phi',H_2$ acquire positive masses squared, see discussion around (\ref{Vu}).

%%%%%%%%%%%%%%%%%%%%%%%%%%%%%%%%%%%%%

 \end{document}